\documentclass[12pt]{article}
\usepackage[english]{babel}
\usepackage{epsfig,amsmath,amsfonts,geometry}[0cm]
\usepackage{mathrsfs,amsmath,amssymb,amsbsy}

\begin{document}
%

\title{\normalsize\bf AN EMERGENT UNIVERSE SUPPORTED BY \\ CHIRAL COSMOLOGICAL FIELDS\\
IN EINSTEIN--GAUSS--BONNET GRAVITY
}

\author{\footnotesize SERGEY V. CHERVON\\
{\vspace{-10pt}\it \scriptsize Astrophysics and Cosmology Research Unit,}\\
{\vspace{-10pt}\it \scriptsize School of Mathematics, Statistics and Computer Science, }\\
{\vspace{-10pt}\it \scriptsize University of KwaZulu-Natal, Private Bag X54 001,}\\
{\vspace{-10pt}\it \scriptsize Durban 4000, South Africa, and} \\
{\vspace{-10pt}\it \scriptsize Laboratory of Gravitation, Cosmology, Astrophysics,}\\
{\vspace{-10pt}\it \scriptsize Ilya Ulyanov State Pedagogical University,}\\
{\vspace{-10pt}\it \scriptsize 100 years V.I.Lenin's Birthday Square, 4, Ulyanovsk 432700, Russia,}\\
{\vspace{-10pt}\it \scriptsize and}\\
{\vspace{-10pt}\it \scriptsize Physics Department, Institute of Physics and Technology,}\\
{\vspace{-10pt}\it \scriptsize Kant Baltic Federal University, A. Nevsky Str.,B. 14,}\\
{\vspace{-10pt}\it \scriptsize Kaliningrad 236041, Russia,}\\
{\vspace{-10pt}\it \scriptsize chervon.sergey@gmail.com \vspace{12pt}}\\
{\footnotesize SUNIL D. MAHARAJ}\\
{\vspace{-10pt}\it \scriptsize Astrophysics and Cosmology Research Unit,} \\
{\vspace{-10pt}\it \scriptsize School of Mathematics, Statistics and Computer Science,}\\
{\vspace{-10pt}\it \scriptsize  University of KwaZulu-Natal, Private Bag X54 001} \\
{\vspace{-10pt}\it \scriptsize  Durban 4000, South Africa} \\
{\vspace{-10pt}\it \scriptsize maharaj@ukzn.ac.za \vspace{12pt}}\\
{\footnotesize AROONKUMAR BEESHAM}\\
{\vspace{-10pt}\it \scriptsize Department of Mathematical Sciences,  University of Zululand, }\\
{\vspace{-10pt}\it \scriptsize Private Bag X1001, Kwa-Dlangezwa 3886, South Africa }\\
{\vspace{-10pt}\it \scriptsize beeshama@unizulu.ac.za \vspace{12pt}}\\
{\footnotesize ALEKSANDR S. KUBASOV}\\
{\vspace{-10pt}\it \scriptsize Laboratory of Gravitation, Cosmology, Astrophysics,}\\
{\vspace{-10pt}\it \scriptsize Ilya Ulyanov State Pedagogical University,}\\
{\vspace{-10pt}\it \scriptsize 100 years V.I. Lenin's Birthday Square, 4, Ulyanovsk 432700, Russia,}\\
{\vspace{-10pt}\it \scriptsize as-kubasov@rambler.ru}}
\maketitle
\thispagestyle{empty}

\begin{abstract}

We propose the application of the chiral cosmological model (CCM) for the Einstein--Gauss--Bonnet (EGB) theory 
of gravitation with the aim of finding new models of the Emergent Universe (EmU) scenario. We analysed the 
EmU supported by two chiral cosmological fields for a spatially flat universe, while we have used three chiral fields 
when we investigated open and closed universes. To prove the validity of the EmU scenario we fixed the scale factor 
and found the exact solution by decomposition of EGB equations and solving the chiral field dynamics equation.
 To this end, we suggested the decomposition of the EGB equations in such a way that the first chiral field is 
 responsible for the Einstein part of the model, while the second field, together with kinetic interaction term, is connected with the
 Gauss--Bonnet part of the theory. We proved that both fields are phantom ones under this decomposition, 
 and that the model has a  solution if the kinetic interaction between the fields equals a constant. 
 We have presented the exact solution in terms of cosmic time. This was done for a spatially flat universe. 
 In the case of open and closed universes we introduced the third chiral field (canonical for closed and 
 phantom for open universe) which is responsible for the EGB and curvature parts. The solution of the 
 third field equation is obtained in quadratures.
 Thus we have proved that the CCM is able to support EmU scenario in EGB gravity for spatially 
 flat, open and closed universes.

\end{abstract}


\section{\bf\rm Introduction}
The discovery of the acceleration in the expanding Universe  at the end of twentieth century led to intensive investigations of modified gravity and 
field theories \cite{clifton2011}. Theories such as scalar-tensor gravitation (which may be considered as Einstein's gravitational field coupled to a self-interacting scalar 
field in the Einstein frame), $f(R)$ gravitation, Lovelock gravity, and its special variant of $5D-6D$ gravity as Einstein-€"Gauss-Bonnet (EGB) 
gravitation, can be related to these theories. We also mention the generalization of EGB theory to $f(R_{GB})$ gravity which acts in $4D$ spacetime
\cite{odintsov14}. 

On the other hand new data from astrophysical observations obtained from the cosmic project WMAP, 
Planck and BICEP2 provide  possibilities to determine restrictions on the parameters of a basic theory according to the following scheme. 
We may start from investigations of gravitational dynamics and finding background solutions. 
Then we develop cosmological perturbation theory, and solve the perturbed equations,
 usually in the long-wave and short-wave approximations. 
 Afterwards it is necessary to  find a power spectrum and spectral indexes which together with the 
 tensor-to-scalar ratio may be used to confront observational digital values. 

Our study belongs to the investigation of multidimensional gravity actively developed
 by V. N. Melnikov and  his group  (for a review, see \cite{melnikov02}). In the present 
 work we make the first step in building a model in this scenario, namely we are searching for exact solutions 
 for the background equations in $5D$ EGB gravity for the Emergent Universe (EmU). Connection to the real $4D$ 
 cosmology may be performed in several ways. One  way is the consideration of the results in this article 
  as  bulk-solutions in brane cosmology.
  The corresponding $4D$ cosmological solutions may be
  obtained by the superpotential method, which is related to developments in brane cosmology  \cite{astashenok2013}.  
 Another possibility to reach $4D$ FRW cosmology is to perform a compactification procedure of $5D$ 
 FRW cosmology, and to prove that chiral fields belong to the brane. Also we may suggest the relevance 
 of our obtained solutions to $f(R_{GB})$ gravity which  is known to act in $4D$ spacetime \cite{odintsov14}. 

The idea of an EmU was proposed by Ellis and Maartens \cite{ellmaa02}. This scenario 
avoids the initial singularity and there is no need for a quantum era. The universe in the infinite 
past is in an almost static state, and then evolves into an inflationary era. Mukherjee {\it et al} \cite{Mukherjee06} showed that it is possible to have a spatially flat emergent universe if one allows for the existence of a perfect fluid together with some type of exotic matter.

The article of Beesham {\it et al} \cite{bechma09} gives the motivation for the study of the EmU 
 scenario within the context of chiral fields in the framework of the nonlinear sigma model. 
 Beesham {\it et al} \cite{bcmk13qm} found solutions with phantom and 
 canonical scalar fields, including a solution valid for all time.
More recently, Beesham {\it et al} \cite{bcnk13b} presented new classes of exact solutions, 
discussed the potential and kinetic interaction of the chiral fields and calculated key cosmological parameters.

In the work \cite{mukcha10} the authors have studied a general approach  to the EmU in EGB cosmology. 
Their discussion was largely qualitative, focusing on the possibility of the existence of solutions and corresponding restrictions.
Some restrictions for the model with canonical and phantom scalar fields coupled to a perfect fluid with a linear equation of state 
have been obtained. A similar analysis was done for tachyon and phantom tachyon fields. 
However, in \cite{mukcha10}, the authors did not really find  exact solutions, 
to  confirm the existence of the EmU scenario in FRW cosmology derived from EGB gravity. 
In the present article we show the possible development of EmU scenario 
by virtue of the exact solutions in EGB cosmology in the framework of a chiral cosmological model. 

In the present work, we study the EmU universe supported by two chiral fields within the context of $5D$ EGB theory. 
Investigation of theories of higher dimensional gravitation is dictated by novel
 results in high energy physics, and particularly in string theory \cite{clifton2011}.  
 The classical analogue of the effective string theory is  represented by the low energy effective 
 action, containing quadratic or higher powers of curvature terms. Similar terms with 
 high power in curvature appear after the renormalization procedure of 
 quantum field theory in the curved space background. 
For these theories the field equations become  fourth order and ghosts appear.
 To avoid these difficulties Lovelock suggested a special combination of high order terms which lead to the 
 field equations of the second order. This means that ghost terms will not appear. 

The paper is organised as follows: In section 2, we present the equations for the 
chiral cosmological model in EGB gravity. In section 3, we specialise to the spatially flat EmU; 
we present the decomposition of EGB gravity equations  and solve the dynamical equations. 
Section 4 deals with closed and open EmU with three chiral cosmological fields; 
decomposition for this case and solution for the third field equation are obtained in quadratures. 
Finally our discussion follows.

\section{\bf\rm The Chiral Cosmological Model in EGB Gravity}

We will follow the standard prescription for EGB gravity \cite{clifton2011} considering the EGB action
\begin{equation}\label{egb-actn}
S=\frac{1}{2}\int d^5x\sqrt{-g}\left( R+ \alpha_{GB} R_{GB}\right) + S_m
\end{equation}
with a matter part $S_m$ as an action of the CCM in 5 dimensions
\begin{equation}\label{ccm-actn}
S_m=S_{ccm}=\int d^5x\sqrt{-g}\left[\frac{1}{2} h_{AB}(\varphi)\varphi^A_{,a}\varphi^B_{,b}g^{ab}-V(\varphi)\right].
\end{equation}
We set here the 5-dimensional Einstein gravitational constant $\kappa=1$. 
The indices $a,b, \dots$ take on values $0,1,2,3,5 $ and $\alpha_{GB}$ is the GB coupling parameter. 
The notation for the CCM corresponds to those in the work \cite{bechma09}.
The Gauss--Bonnet term is
\begin{equation}\label{gb-term}
R_{GB}=R^2-4R_{ab}R^{ab}+R_{abcd}R^{abcd}.
\end{equation}

The variation of the action (\ref{egb-actn}) with respect to metric tensor gives the equations of EGB gravitation
\begin{equation}\label{egb-grav}
G_{ab}-\alpha_{GB} H_{ab}=T_{ab},
\end{equation}
where $ G_{ab} = R_{ab}-\frac{1}{2} g_{ab}R $ is the Einstein tensor, and $H_{ab}$ is the Lovelock tensor defined by
\begin{equation}\label{lovlk-tens}
H_{ab}=4R_{ac}R^c_b+4R^{cd}R_{acbd}-2RR_{ab}-2R^{cde}_a R_{bcde} +\frac{1}{2} g_{ab}R_{GB}.
\end{equation}
The chiral cosmological field equations may be derived by varying the action (\ref{ccm-actn})
with respect to scalar fields $ \varphi^C $
\begin{equation}\label{chi-fields}
\frac{1}{\sqrt{-g}}\partial_a\left(\sqrt{-g} \varphi^{,a}_A\right) -
\frac{1}{2} \frac{\partial h_{BC}}{\partial \varphi^A}\varphi^C_{,a}\varphi^B_{,b} g^{ab}+V_{,A}=0,
\end{equation}
where $V_{,A}=\frac{\partial V}{\partial \varphi^A} $.
It should be noted that in the present article all considerations are applied to a 5-dimensional spacetime.

Let us turn our attention to cosmology. We take the metric of the 5-dimensional Friedmann--Robertson--Walker (FRW) universe in the form
\begin{equation}\label{frw-5d}
dS^2=-dt^2+a(t)^2\left(\frac{dr^2}{1-\epsilon r^2} + r^2 (d\theta + \sin^2 \theta (d\varphi^2 + \sin^2 \varphi d\chi^2) )\right).
\end{equation}
Here $\epsilon = -1, 0, +1$ for open, spatially flat and closed universes respectively.

If we take only one effective scalar field as in earlier treatments
\cite{bcmk13qm,chervon-qm13} we may introduce a scalar field multiplett (CCM) which will have the 
same gravitational properties but different field dynamics. For EGB gravity, the situation 
is rather different as we will demonstrate later on in this article. 
We can state that the more general features of the CCM are exhibited in the two component model. 
Thus we will consider the two component CCM with the chiral (target space) metric
\begin{equation}\label{ts-metric}
ds_{ts}=h_{11}d\phi^2 + h_{22}(\phi, \psi)d\psi^2 .
\end{equation}
The energy momentum tensor of the target space (\ref{ts-metric}) in the $5D$ spacetime (\ref{frw-5d}) takes the form
\begin{equation}\label{ccm-emt}
T_{ab}=h_{11}\phi_{,a}\phi_{,b}+h_{22}\psi_{,a}\psi_{,b}-g_{ab}\left[\frac{1}{2} h_{11} \phi_{,c}\phi^{,c} +\frac{1}{2} h_{22}\psi_{,c}\psi^{,c}-V(\phi, \psi) \right],~~
\end{equation}
where $h_{11}=constant. $

 We can now present the basic equations of the CCM (\ref{chi-fields}) and 
 EGB gravity (\ref{egb-grav}) in $5D$ FRW spacetime (\ref{frw-5d}).
This system of equations takes the form
\begin{eqnarray}
\label{basic-egb1}
&H^2+\frac{\epsilon}{a^2}+\alpha_{GB}\left(H^2+\frac{\epsilon}{a^2}\right)^2=
\frac{1}{6} \left(\frac{1}{2} h_{11}\dot{\phi}^2 + \frac{1}{2} h_{22}(\phi,\psi)\dot{\psi}^2 +V(\phi, \psi)\right),\\
\label{basic-egb2}
&\left[ 1+2\alpha_{GB} \left(H^2+\frac{\epsilon}{a^2}\right) \right] \left(\dot{H}-\frac{\epsilon}{a^2}\right) =-\frac{1}{3}\left(h_{11}\dot{\phi}^2 +  h_{22}(\phi,\psi)\dot{\psi}^2\right), \\
\label{basic-fs1}
&h_{11}\ddot{\phi} +4H h_{11}\dot{\phi}-\frac{1}{2}\frac{\partial h_{22}}{\partial\phi} \dot{\psi}^2 + \frac{\partial V}{\partial \phi}=0,~ h_{11}=constant,\\
\label{basic-fs2}
&h_{22}(\phi,\psi)\ddot{\psi} + \dot{h_{22}}(\phi,\psi)\dot{\psi}+4H h_{22}(\phi,\psi)\dot{\psi}-\frac{1}{2}\frac{\partial h_{22}}{\partial\psi} \dot{\psi}^2+ \frac{\partial V}{\partial \psi}=0.
\end{eqnarray}
Our investigation of these equations will be applied first to a spatially flat EmU, i.e., with the model 
proposed by Mukherjee {\it et al} in the work \cite{Mukherjee06}.

\section{\bf\rm Spatially flat EmU with two chiral cosmological fields}

The basic EGB model equations (\ref{basic-egb1})-(\ref{basic-egb2}) will be simplified taking into account $\epsilon=0$:

 \begin{eqnarray}
 \label{flat-egb1}
 &H^2+\alpha_{GB}H^4=
 \frac{1}{6} \left(\frac{1}{2} h_{11}\dot{\phi}^2 + \frac{1}{2} h_{22}(\phi,\psi)\dot{\psi}^2 +V(\phi, \psi)\right),\\
 \label{flat-egb2}
& \left[ 1+2\alpha_{GB} H^2 \right] \dot{H} =-\frac{1}{3}\left(h_{11}\dot{\phi}^2 + h_{22}(\phi,\psi)\dot{\psi}^2\right).
 \end{eqnarray}
The CCM equations (\ref{basic-fs1})-(\ref{basic-fs2}) do not change.
We extract the potential $V$ from Eqs. (\ref{flat-egb1})-(\ref{flat-egb2})
\begin{equation}\label{V-on-t}
\frac{V}{6}= H^2 +\frac{1}{4} \dot{H} + \alpha_{GB} H^2 (H^2 + \frac{1}{2} \dot{H}).
\end{equation}
It should be mentioned here that Eq. (\ref{flat-egb2}) can be obtained from a 
linear combination of the chiral fields Eqs. (\ref{basic-fs1})$ \dot\phi $+ (\ref{basic-fs2}) $\dot{\psi}$ and using Eq. (\ref{flat-egb1}).

Let us consider the case of a constant potential. In Einstein gravity a constant potential is equivalent 
to a cosmological constant, admitting the exact solution \cite{chervon-qm13}, included in 
inflation scenarios, and may be considered as a useful tool for numerical solutions \cite{abbche13}.

In EGB Gravity Eqs. (\ref{flat-egb1}) and (\ref{flat-egb2}) can be reduced to the form
\begin{equation}\label{egb-V-const}
\dot{H}\left(\alpha_{GB} H^2+\frac{1}{2} \right)+H^2\left(1+\alpha_{GB} \right)=\frac{1}{6}\Lambda
\end{equation}
which gives us an opportunity to perform the integration.
The solution of (\ref{egb-V-const}) can be expressed exactly by the relation
\begin{equation}\label{sol-special}
t-t_*=-\alpha_{GB} (1+\alpha_{GB})\left[H+\frac{1+\alpha_{GB} \frac{1}{3}\Lambda (1+\alpha_{GB})}{4\alpha_{GB}} \ln \left|\frac{H-\sqrt{\frac{1}{6}\Lambda (1+\alpha_{GB})}}{H+\sqrt{\frac{1}{6}\Lambda (1+\alpha_{GB})}} \right|  \right].
\end{equation}

It is clear from Eq. (\ref{sol-special}) that it is impossible to obtain $H$ as a function on $t$ explicitly. If we consider small $\Lambda$ then we will obtain a collapsing universe with $a(t) \propto \exp \left[-\frac{(t-t_*)^2}{2\alpha_{GB} (1+\alpha_{GB})}\right]$ due to the EGB gravity influence. Taking $\alpha_{GB} =0 $ in (\ref{egb-V-const}) we obtain the solution
\begin{equation}
H=\sqrt{\frac{\Lambda}{6}}\tanh \left(\sqrt{\frac{2\Lambda}{3}} t\right),~~a(t)=a_s\left(\cosh\sqrt{\frac{2\Lambda}{3}}t\right)^{1/2}.
\end{equation}
Observe that the solution (\ref{sol-special}) is different from the 4-dimensional result of Chervon \cite{chervon-qm13}; he obtained the forms
$
H =\sqrt{\frac{\Lambda}{3}}\tanh (\sqrt{3\Lambda}t),~~
a=a_*[\cosh (\sqrt{3\Lambda}t)]^{1/3}
$
which are not the same as in the 5-dimensional result (\ref{sol-special}).

\subsection{\bf\it Decomposition of EGB system}

The scale factor of the EmU we choose in the most general form \cite{Mukherjee06,bcmk13qm}
\begin{equation}\label{a_emu}
a(t)=A\left(\beta + e^{\alpha t}\right)^m.
\end{equation}

Following the method of Beesham {\it et al} \cite{bcnk13b} we decompose the EGB and chiral fields equations in the following way:
\begin{eqnarray}
\label{dec-phi}
&\dot{H}=-\frac{2}{3}K_{(\phi)}, ~K_{(\phi)}=\frac{1}{2} h_{11}\dot{\phi}^2,\\
\label{dec-psi}
&2\alpha_{GB} H^2 \dot{H}=-\frac{2}{3}K_{(\psi)}, ~K_{(\psi)}=\frac{1}{2} h_{22}(\phi,\psi) \dot{\psi}^2,\\
\label{V-phi}
&H^2+\frac{1}{4} \dot{H} =\frac{1}{6}V_{(\phi)}, \\
\label{V-psi}
&H^4 +\frac{1}{2} H^2 \dot{H}=\frac{1}{6\alpha_{GB}}V_{(\psi)}.
\end{eqnarray}
From (\ref{dec-phi}) and (\ref{dec-psi}) one can see that both fields should be phantom 
ones for the EmU, because $\dot{H}>0$. Thus we can set $h_{11}=-1$ and transform $h_{22} \rightarrow -h_{22}$.

Eq. (\ref{dec-phi}) has the solution
\begin{equation}\label{sol-phi}
\phi-\phi_i= 2\sqrt{3m}\arctan \left(\frac{e^{\alpha t/2}}{\sqrt{\beta}}\right)
\end{equation}
where $\phi_i $ corresponds to the value at $t=-\infty $.
Using the solution (\ref{sol-phi}) above we can reconstruct the potential $V_{(\phi)}$. The result is
\begin{equation}\label{sol-V-phi}
 V_{(\phi)} =\frac{3}{2}m\alpha^2 \sin^2 2\tilde{\phi} \left(m \tan^2\tilde{\phi} +\frac{1}{4} \right),
\end{equation}
where $\tilde{\phi}=\frac{\phi-\phi_i}{2\sqrt{3m}} $.

The formulae in terms of cosmic time $t$ for the potential and kinetic energy of $\psi$ read
\begin{eqnarray}
\label{V-psi-t}
&V_{(\psi)}=6 \alpha_{GB} m^3 \alpha^4 e^{3\alpha t}\frac{m e^{\alpha t}+\frac{1}{2} \beta}{(\beta +e^{\alpha t})^4},\\
\label{K-psi-t}
&\frac{1}{2} h_{22}(\phi,\psi)\dot{\psi}^2=3\alpha_{GB} m^3 \alpha^4 \beta \frac{e^{3\alpha t}}{(\beta +e^{\alpha t})^4}.
\end{eqnarray}
Note that with this result the EGB Eqs. (\ref{flat-egb1}) and (\ref{flat-egb2})
are satisfied. Our task now is to show that the dynamics of the chiral fields will not contradict the solutions of the gravitational equations.

\subsection{\bf\it The solution of dynamic equations}

Because of the negative sign in (\ref{dec-phi}) and (\ref{dec-psi}), we have changed the sign of $h_{22}$ and set $h_{11}=-1$. With this, we emphasize that both chiral fields are phantom.

Direct insertion of the solution (\ref{sol-phi}) into (\ref{basic-fs1}), and using the formulae
\begin{eqnarray}
&H=\frac{m\alpha e^{\alpha t}}{\beta +e^{\alpha t}},\\
&\dot{\phi}=\sqrt{3m\beta}\alpha\frac{e^{\alpha t/2}}{\left(\beta+e^{\alpha t}\right)},\\
&\ddot{\phi}= \sqrt{3m\beta}\frac{\alpha^2}{2}\left[\frac{e^{\alpha t/2}\left(\beta-e^{\alpha t}\right)}{\left(\beta+e^{\alpha t}\right)^2}\right],\\
&\frac{\partial V_{(\phi)}}{\partial \phi}= \frac{\sqrt{3m}\alpha^2}{2}\left[\sin 4\tilde{\phi}\left(m \tan^2 \tilde{\phi}+\frac{1}{4}\right)+4m \sin^2 \tilde{\phi}\tan\tilde{\phi}\right],
\end{eqnarray}
lead to conclusion that the term  $\frac{1}{2} \frac{\partial h_{22}}{\partial \phi}\dot{\psi}^2 $  should be equal to zero. Indeed, making reconstruction of the derivation of the $ \phi $-part of the potential $\frac{\partial V_{(\phi)}}{\partial \phi}$ in terms of cosmic time $t$ we obtain
\begin{equation}
\frac{\partial V_{(\phi)}}{\partial \phi}=\sqrt{3m\beta}\frac{\alpha^2}{2} e^{\alpha t/2}\frac{\left(e^{\alpha t}(8m-1)+\beta\right)}{(\beta +e^{\alpha t})^2},
\end{equation}

Thus the result $\frac{1}{2} \frac{\partial h_{22}}{\partial \phi}\dot{\psi}^2 =0$  means that $ h_{22}$ is the 
function on $\psi$ only. With this we can set $h_{22}=1$ without loss of generality, and look
 for the solution of the second field equation (\ref{basic-fs2}). To this end we will multiply the Eq. (\ref{basic-fs2}) on $\dot{\psi}$, and use the relation
\begin{equation}
\dot{V}_{(\psi)}=\lambda_1\alpha e^{3\alpha t}\frac{\left(e^{\alpha t}(4m-\frac{1}{2})+\frac{3}{2}\beta\right)}{(\beta +e^{\alpha t})^5},
\end{equation}
where $ \lambda_1=6\alpha_{GB} m^3 \alpha^4 \beta $.
With this, Eq. (\ref{basic-fs2}) will be identically satisfied.

So our task is to find the second chiral field $\psi$. This can be done from the decomposition ansatz (\ref{K-psi-t}) where we taking into account $h_{22}=1$.
The solution is
\begin{equation}\label{exact-psi}
\psi-\psi_i=\frac{\sqrt{\lambda_1}}{\alpha}\left[\frac{1}{\sqrt{\beta}}\arctan \left(\frac{e^{\alpha t/2}}{\sqrt{\beta}}\right)-\frac{e^{\alpha t/2}}{\beta+e^{\alpha t}}\right],
\end{equation}
where $\psi_i$ corresponds to the value at $t \rightarrow -\infty $.
It is clear that we could not obtain the dependance of $\psi$ on $t$ explicitly from the solution (\ref{exact-psi}).

Thus the exact solution we obtained are represented by solutions for chiral 
cosmological fields (\ref{sol-phi}) and (\ref{exact-psi}); the potential $V_{(\phi)}$ in $(\ref{sol-V-phi})$ and the potential $V_{(\psi)}$ are defined 
via dependence on cosmic time in (\ref{V-psi-t}).

\section{\bf\rm Closed and open EmU with three chiral cosmological fields}

Now we are ready to take into consideration closed and open EmU. 
Following the developed method we are including, in the chiral cosmological model, the 
third field $\chi $ which is responsible for the curvature of the FRW Universe and EGB gravity. 
With this aim we have to extend the decomposition with the kinetic and potential energies of $\chi$:
\begin{eqnarray}\label{K_chi}
& K_{(\chi)}=\frac{1}{2} h_{33}\dot{\chi}^2=\frac{3}{2}\frac{\epsilon}{a^2}\left(1+
2\alpha_{GB}\left(H^2-\dot{H}+\frac{\epsilon}{a^2}\right)\right)\\
\label{V_chi}
&V_{(\chi)}=3\alpha_{GB}\frac{\epsilon^2}{a^4}+\frac{\epsilon}{a^2}\left(\frac{9}{2}+
3\alpha_{GB}(\dot{H}+3H^2\right)
\end{eqnarray}

The equation on the chiral field $\chi$ for the constant kinetic interaction $h_{33}=constant $ is
\begin{equation}\label{chi-field}
h_{33}\ddot{\chi}+4Hh_{33}+\frac{\partial V_{(\chi)}}{\partial \chi}=0.
\end{equation}
Multiplying  Eq. (\ref{chi-field}) by $\dot{\chi}$, and inserting $\dot{\chi},\ddot{\chi}$ 
from Eq. (\ref{K_chi}), and $\dot{V}_{(\chi)}$ from Eq. (\ref{V_chi}), we can check that by consequence
Eq. (\ref{chi-field}) is satisfied, and the very equation is satisfied too if $\dot{\chi} \neq 0$.

Now our task is to define the third field $\chi $ from Eq. (\ref{K_chi}).
Let us note here that for an open universe ($ \epsilon=-1$) the sign of $h_{33}$ should be negative, 
i.e., the third field $\chi$ should be phantom. For a closed universe ($ \epsilon=1$) the third field $\chi$ should be 
canonical. With this remark we can write the general formula for $\chi$:
\begin{equation}\label{chi_int}
\chi= \frac{\sqrt{3}}{A}\int dt ~(\beta +e^{\alpha t})^{-m}\sqrt{1
+2\alpha_{GB}\left(\frac{m\alpha^2e^{\alpha t}(me^{\alpha t}-\beta)}{(\beta+e^{\alpha t})^2}+\frac{\epsilon}{A^2(\beta+e^{\alpha t})^{2m}}\right)}
\end{equation}

Thus we have obtained the solution for open and closed universes with three chiral cosmological fields. 
It is difficult to perform the integration for the third chiral field $ \chi $ in Eq.(\ref{chi_int}). 
Therefore we turn our attention to the case when $\beta \approx 0 $. This situation studied by Paul {\it et al}  \cite{paul2010}. 
Note that we analyze the integration only for $ \chi $ in Eq. (\ref{chi_int}), not for the other equation.

Taking into account $\beta \approx 0 $ in Eq. (\ref{chi_int}) we obtain the general formula
\begin{equation}\label{chi_approx}
\chi= \frac{\sqrt{3}}{A}\int dt ~e^{-2m\alpha t}
\sqrt{e^{2m\alpha t}+2\alpha_{GB}\left(m^2\alpha^2e^{2m\alpha t} +\frac{\epsilon}{A^2}\right)}.
\end{equation}
To simplify the expression (\ref{chi_approx}) let us introduce a new variable and constants. Let
$$
Z=e^{2m\alpha t},~~B^2=1+2\alpha_{GB} m^2 \alpha^2,~~D^2=\frac{2\alpha_{GB}}{A^2B^2}.
$$
We have two possible solutions
\begin{eqnarray}
&\epsilon =+1,~~h_{33}=+1,\\
&\chi -\chi_*=\pm \frac{\sqrt{3}B}{Am\alpha}\left[-\frac{1}{2}\frac{\sqrt{Z+D^2}}{Z}+
\frac{1}{4D}\ln\left(\frac{\sqrt{Z+D^2}-D}{\sqrt{Z+D^2}+D}\right)\right],\\
&\epsilon =-1,~~h_{33}=-1,\\
&\chi -\chi_*=\pm \frac{\sqrt{3}B}{Am\alpha}\left[-\frac{1}{2}\frac{\sqrt{Z-D^2}}{Z}+
\frac{1}{4D}\arctan\left(\frac{\sqrt{Z-D^2}}{D}\right)\right],~~Z\geq D^2.
\end{eqnarray}
The inequality  $ Z\geq D^2 $ means that the solution will be valid only for time
$$
t \geq \frac{\ln \left(2\alpha_{GB} A^{-2}\alpha^{-2}\right)}{2m\alpha}.
$$
When $t \rightarrow \infty$ the third chiral field tends to some constant.

Let us mention that the case $ \epsilon =+1,~~h_{33}=
-1$ also admits a solution, while the case $\epsilon =-1,~~h_{33}=+1$ 
may be valid only for some period of time.

\section{\bf\rm Discussion}

Our study is motivated by the great attention paid recently to modified gravity theories due to the possible "gravitational" 
explanation of the acceleration in the expanding Universe. Amongst modified gravity theories, the multidimensional 
one plays an important role both from the view of mathematical physics  and from the connection to 
realistic $4D$ gravitation under compactification procedure of additional dimensions \cite{melnikov02}. 
On the other hand an emergent universe, as well as eternal inflation and a universe with cyclic  evolution
 have many criticisms concerning stability and singularity. In our consideration we are not concerned with these problems 
 because of the possibility to restrict our obtained results to positive time. Consideration of negative time is only of a
  purely mathematical character. The methods applied to the EmU scenario are also 
  applicable to exponential, power-law and others kinds of early inflation.
  
In this article we proposed application of a chiral cosmological model in EGB gravity. 
Considering the two-component CCM with two phantom fields in the spatially flat FRW 
universe we proved the existence of an exact solution and found it. In the case of closed and 
open universe we introduced the third field and found the exact solution in quadratures. 
Thus we strongly confirm the existence of the EmU scenario in EGB gravity for spatially flat, open and closed universes.

\section{\bf\rm Acknowledgments}

SVC is thankful to the University of KwaZulu-Natal, the University
of Zululand and  the NRF for financial support and warm
hospitality during his visit in 2013 to South Africa. SDM
acknowledges that this work is based upon research supported by
the South African Research Chair Initiative of the Department of
Science and Technology and the National Research Foundation. SVC and ASK
would like to mention that the part of this work was carried out within
the framework of a State Order of the Ministry of Education and Science
of the Russian Federation in accordance with Project No.2014/391.

\end{document}